\documentclass{elsart}
\usepackage{graphics}
\begin{document}
\begin{frontmatter}
\title{The Foam Analogy: From Phases to Elasticity}
\author{William Kung}
\address{Department of Physics and Astronomy, University of
Pennsylvania, Philadelphia, PA 19104-6396}
\author{P. Ziherl}
\address{Department of Physics, University of Ljubljana, SI-1000 Ljubljana,
Slovenia\\
J. Stefan Institute, Jamova 39, SI-1000 Ljubljana, Slovenia}
\author{Randall D. Kamien}
\address{Department of Physics and Astronomy, University of
Pennsylvania, Philadelphia, PA 19104-6396}
\date{\today}
\begin{abstract}
By mapping the interactions of colloidal particles onto the
problem of minimizing areas, the physics of foams can be used to
understand the phase diagrams of both charged and fuzzy colloids.
We extend this analogy to study the elastic properties of such
colloidal crystals and consider the face-centered cubic,
body-centered cubic and A15 lattices. We discuss two types of soft
interparticle potentials corresponding to charged and fuzzy
colloids, respectively, and we analyze the dependence of the
elastic constants on density as well as on the parameters of the
potential. We show that the bulk moduli of the three lattices are
generally quite similar, and that the shear moduli of the two
non-close-packed lattices are considerably smaller than in the
face-centered cubic lattice. We find that in charged colloids, the
elastic constants are the largest at a finite screening length,
and we discuss a shear instability of the A15 lattice.
\end{abstract}
\end{frontmatter}

\section{Introduction}

Soft materials are often subjugated into classes: ``colloids,
polymers, or emulsions", or ``liquid crystals,  membranes, or
foams".  While these systems all share common mechanical and
energetic scales, each subfield has developed its own set of
theoretical tools, often required and inspired by real-world
applications.  Recently~\cite{Ziherl00,Primoz01,Kung02} we have combined
ideas from these subfields to
study colloids which, in addition to the standard
excluded-volume interactions, softly repel each other, either
through a surface polymer coating (``fuzzy'')~\cite{Whetten99} or
via a screened Coulomb interaction~\cite{Sirota89}.  While
free-volume theory provides remarkably accurate accounting of the
free energy of packing the hard cores~\cite{Curtin87}, additional
pair interactions are hard to implement analytically to determine
in detail the phase behavior of these systems.  We have argued
that soft repulsion favors larger separations between the colloid
surfaces which, in turn, implies that the area of the
(mathematical) interface partitioning the Voronoi cells of each
colloid should be a minimum~\cite{Ziherl00}. Employing this idea
and borrowing from the physics of foams, we have successfully
understood the phase diagrams of these and other systems with
similar morphologies, {\sl e.g.} dendritic
polymers~\cite{Balagurusamy97}, diblock
copolymers~\cite{McConnell96} and star polymers~\cite{gidomacro}.
In this paper we extend this analogy in order to determine the
elastic properties of these materials when assembled into three
lattices of interest: the face-centered cubic lattice (fcc), which
maximizes the packing density of hard spheres~\cite{Hales00}, the
body-centered cubic lattice (bcc), which minimizes the overlap
volume of spheres which cover space~\cite{Sloane}, and the A15
lattice, which is conjectured to minimize the interfacial area of
the Voronoi cells~\cite{Weaire94}.

Fortunately, there exists a considerable body of experimental data and facts
on many of these colloidal systems: from crystallographic
data~\cite{Russelbook} to charged systems~\cite{Hone83,Kremer86,Rosenberg87,Rascon97} and lyotropic systems~\cite{Balagurusamy97,Percec98,Hudson99}. The abundance of data
can be attributed in part to the relative ease of manipulating
colloidal systems by chemical means in the laboratory, and the
resulting systems can be readily analyzed using optical
techniques~\cite{Russelbook,Hachisu73,Yoshiyama86,Monovoukas89}.
In fact, colloidal systems would serve as an ideal experimental
playground provided that a recipe exists that would enable predictions of
their external mechanical, thermal and electrical properties as
functions of internal variables such as the size of colloidal
particles and the form of interparticle interaction.

Prior efforts have been made to understand these colloidal systems
from first principles. The more traditional routes include
molecular dynamics simulations~\cite{Robbins88} or analytical
methods like the density functional approach~\cite{density91}.
While these tools have proven important in our current
understanding, they do not provide a robust explanation of the
stability of colloidal crystals.  Our analogy with the physics of
foams provides a unifying explanation of non-close-packed
structures in fuzzy colloids and charged polystyrene spheres.   In
both of these systems the interaction potential consists of a hard
core decorated with a ``soft", purely repulsive part. In
dendrimers, for instance, the hard core is composed of the
aromatic internal part of the micelle-forming conical
macromolecules, and the soft corona comes from the interdigitated
alkyl chains of the outer part of the
micelles~\cite{Balagurusamy97}. The charged colloids, on the other
hand, consist of polystyrene spheres and the soft repulsive tail
is their short-range, screened Coulomb potential.

In what follows, we summarize the main concepts behind the foam
analogy in Section II. We see that geometry provides a natural
setting to relate the  microscopic properties of colloids such as
lattice spacing or interparticle interaction to phase diagrams
delineating macroscopic and thermal properties of these systems.
In section III, we extend this foam analogy to
study the elastic properties of these materials by considering small
perturbations to the ``foams''.  Section IV concludes the paper.

\section{The Foam Analogy}

In this section we sketch the salient features of the mapping between fuzzy colloids and
foams.  When studying the interactions of spheres, one limit is to treat them as point
particles with infinite curvature.  Our approach is the other extreme: we treat the interactions
between the spheres as if they are between flat, zero-curvature surfaces.

If the system contains only one type of particle, we may divide
the volume into equal volume cells, each singly occupied. At fixed
density, the volume of the system is the volume of the hard cores
and the excess volume of either the salt solution in charged
systems or the soft chains within the corona in fuzzy colloids.
Since this latter volume can be viewed as enveloping the
individual spheres, we imagine breaking the volume up into a
lattice of Voronoi cells, each of which contains a colloidal
particle. The excess volume can then be written, by definition, as
the product of the area of these dividing surfaces and their
average thickness.  Since the volume of the hard cores and the
volume of the whole system is fixed at a given density, the excess
volume is also constant and so:
\begin{equation}
A_Md={\rm constant},
\label{Ad}
\end{equation}
where $A_M$ is total area of these bilayers and $d$ is their average
thickness. The excess volume depends on the particle density of
the system. Using simple geometrical considerations, the volume
per particle subtracting the volume of each particle is
\begin{equation}
A_{\rm \scriptscriptstyle
M}d=2\left(\frac{1}{n}-\frac{\pi}{6}\right) \sigma^{3},
\label{area}
\end{equation}
where $A_{\rm \scriptscriptstyle M}=\gamma^{\scriptscriptstyle X}
\sigma^2n^{-2/3}$, depends on the lattice type
($\gamma^{\rm{fcc}}=5.345$, $\gamma^{\rm{bcc}}=5.308$,
$\gamma^{\rm{A15}}=5.288$) and is a dimensionless quantity
characterizing the magnitude of area.

Given the repulsive nature of both types of systems, the surfaces
of the hard cores within each Voronoi cell would like to be as far
apart from one another as possible. Maximizing $d$ to reduce the
repulsive interaction amounts to minimizing $A_{\rm\scriptscriptstyle M}$ and we are thus
led to consider Kelvin's problem~\cite{Thomson87} of partitioning
space into equal volume cells having the smalled interfacial area.
It follows that the soft repulsion favors the bcc lattice over the
fcc lattice~\cite{Thomson87} and the A15 lattice over
both~\cite{Weaire94,Ziherl00,Balagurusamy97}. The total free
energy of the system can be divided into two contributions: a bulk
part due to the positional entropy of the hard cores in the
Voronoi cells, and an interfacial part due to the surface
interactions between neighboring cells. Other than its heuristic nature,
the main advantage of our
geometrical approach lies in the fact that although it
incorporates only nearest-neighbor effects, our calculations
have a many-body character since both maximization of the packing
density and minimization of the interfacial area are global rather
than local problems.

To compute the bulk free energy of the system, we employ
cellular free-volume theory~\cite{Hill56,Barker63}. In this
approximation, each particle is confined to a cage formed by its
neighbors. The free volume available to each particle's center of
mass is the volume of the Wigner-Seitz cell after a layer of
thickness $\sigma/2$ (where $\sigma$ is the hard-core diameter of
particles) is peeled off of its faces. Since the volume of the
Wigner-Seitz cell depends only on the symmetry of the
lattice, the bulk free energy encodes information on the geometry
of the system. It turns out that the free volume theory yields
good quantitative agreement with available numerical
simulations at high densities in spite of its mean-field nature~\cite{Curtin87}.
At lower densities, as long as the shear elastic
constants are non-zero, we expect that an ``Einstein-crystal''
description of the phonon modes should be adequate. Indeed, for
the systems in which we are interested, the appropriate moduli are
all non-zero in the density regime we are
probing~\cite{Robbinspers}: therefore there should be no ``soft
modes'' which might otherwise contribute strongly to collective
effects.

The use of the free volume theory results in the following
expression for the bulk free energy of the fcc or bcc lattice:
\begin{equation}
F^{\scriptscriptstyle X}_{\rm bulk}=-k_{\scriptscriptstyle
B}T\ln\bigg(\alpha^{\scriptscriptstyle X}\left(\beta^{\scriptscriptstyle X}n^{-1/3}-1\right)^3\bigg),
\label{Fbulk}
\end{equation}
where $n=\rho\sigma^3$ is the reduced density and $\sigma$ is the
hard-core diameter of each colloidal particle. The coefficients
$\alpha^{\scriptscriptstyle \rm{fcc}}=2^{5/2}$ and
$\alpha^{\scriptscriptstyle \rm{bcc}}=2^23^{1/2}$ depend on the
shape of the cells, whereas $\beta^{\scriptscriptstyle\rm{
fcc}}=2^{1/6}$ and $\beta^{\scriptscriptstyle\rm{
bcc}}=2^{-2/3}3^{1/2}$ are determined by their size.

The bulk free energy expression for the A15 lattice is slightly
more complicated but still amenable to an approximate analytical
form~\cite{Ziherl00,Primoz01}:
\begin{eqnarray}
F_{\rm bulk}^{\rm A15}&=&-k_{\scriptscriptstyle
B}T\left[\frac{1}{4}\ln \Bigg(\frac{4\pi
S}{3}\left(\frac{\sqrt{5}}{2n^{1/3}}-1\right)^3\Bigg) \right.
\nonumber \\&+&\left.\frac{3}{4}\ln\Bigg(2\pi
C\left(\frac{\sqrt{5}}{2n^{1/3}}-1
\right)^2\left(\frac{1}{n^{1/3}}-1\right)\Bigg)\right].
\label{Fa15}
\end{eqnarray}
This formula best agrees with the numerical results obtained
within the cellular theory for $S=1.638$ and $C=1.381$.

A comparison of Eqs.~(\ref{Fbulk}) and (\ref{Fa15}) shows that the
free volume theory predicts that at reduced densities below about
$n\approx0.48$, the A15 lattice of hard spheres should become more
stable than the fcc lattice, which is physically inadmissible.
However, this is not essential since a pure hard-core system melts
at reduced densities below $n\approx1$. We note that this anomaly
is not an artifact of the approximations behind Eq.~(\ref{Fa15});
instead, it provides us with a conservative estimate for the range
of validity of the cellular theory in this system.

The interfacial free energy of the system, on the other hand,
incorporates the specific nature of interparticle interactions at the
microscopic level. For charged systems, the interaction is the
screened Coulomb potential. As in Ref.~\cite{Kung02}, the
Debye-H\"uckel approximation and a Derjaguin-like approximation are
both used to obtain the interfacial free energy:
\begin{equation}
F_{\rm c}=64A_{\rm M}k_{\scriptscriptstyle B}Tn_{\rm
b}\kappa^{-1}{\rm tanh}^2\left({1\over4}\Psi_{\rm
s}\right)\exp\left(-\kappa d\right).
\label{Finterfacial}
\end{equation}
where $n_{\rm b}$ is the bulk counterion number density,
$\Psi_{\rm s}$ is the dimensionless surface potential of colloids,
$n$ is the colloid density, and the Debye screening length is
$\kappa^{-1}=\sqrt{\epsilon\epsilon_0 k_{\scriptscriptstyle B}
T/2e^{2}Z^{2}n_{b}}$, itself a function of these control variables
(where $\epsilon\epsilon_0$ is the dielectric constant).

For fuzzy colloids, an argument based on Flory theory of the highly
compressed polymer brushes~\cite{Ziherl00,Primoz01} yields:
\begin{equation}
F_{\rm surf}=\frac{\ell N_0k_{\scriptscriptstyle
B}T}{h}=\frac{2\ell N_0k_{\scriptscriptstyle B}T}{d},
\label{Ffuzzy}
\end{equation}
where $\ell$ is a parameter with the dimension of length that
determines the strength of repulsion, $N_0$ is the number of alkyl
chains per micelle, and $h$, the thickness of the corona, is half
the average thickness of the interdigitated matrix of the chains,
$d$. In both systems, maximization of surface spacing $d$ implies
the minimization of the interfacial area $A_{\rm M}$. From Eq.
(\ref{area}), the density of particles enter directly into the
energies of the system through the minimal-area constraint.
However, the functional dependence on the particle density differs
between them: for the charged colloids, we have a short-range
exponential interaction while for the fuzzy colloids we have an
algebraic interaction, valid at ranges smaller than the thickness
of an uncompressed coronal brush.

The bulk and interfacial free energies together express the
thermodynamics of colloidal systems in a geometrical language. The
formalism directly establishes relations between microscopic
parameters governing the dynamics of the colloidal particles and
the stability of various macroscopic phases with respect to
temperature~\cite{Kung02, Primoz01}. One can conceivably adapt
Eqs.~(\ref{Fbulk}), (\ref{Fa15}), (\ref{Finterfacial}), and
(\ref{Ffuzzy}) to other kinds of lattice geometry, and the foam
analogy can correspondingly be applied to studying the various
solid phases of colloidal systems.  In the following section we
will consider lattices which can be attained through shearing the
fcc, bcc and A15 lattices.

\begin{figure}
\vspace{2mm} \hspace{-5mm}
\includegraphics{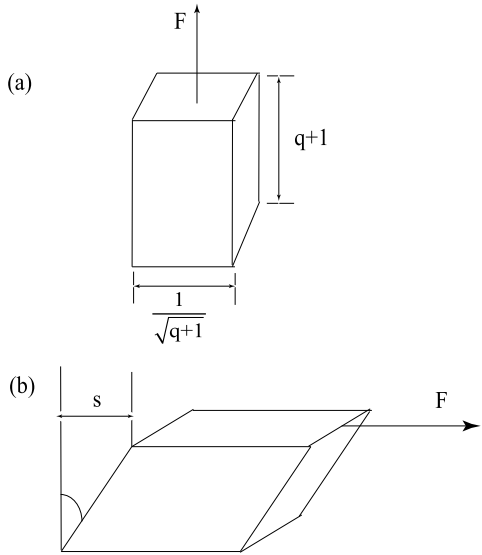}
\vspace*{5mm} \caption{The two different types of deformations:
(a) elongational shear mode; shear deformation along the four-fold
axis (b) simple shear mode; shear deformation along the face
diagonal.  In both cases, we only consider deformations which
preserve the total volume of the unit cell.  The elongational
shear mode is parametrized by $q$, while the simple shear mode is
parametrized by $s$.} \label{deformations}
\end{figure}
\begin{figure}
\vspace{2mm} \hspace{-6mm}
\includegraphics{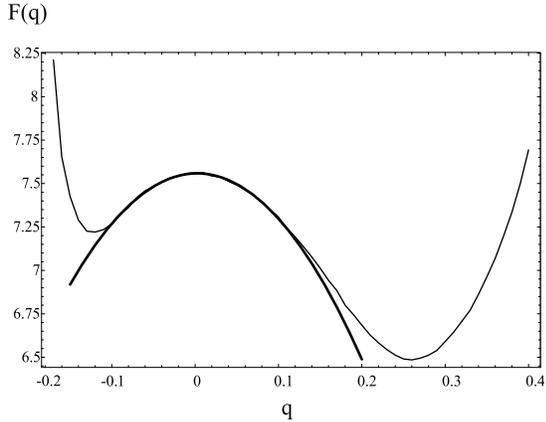}
\caption{The bulk free energy of the bcc lattice as a function of
the elongational shear at $n=1.1$. The thick line represents the
quadratic fit to the curve at $q=0$ which determines the bulk part
of the shear modulus. The minimum at $q\approx0.26$ corresponds to
the fcc lattice.} \label{qFbcc}
\end{figure}
\begin{figure}
\vspace{2mm} \hspace{-6mm}
\includegraphics{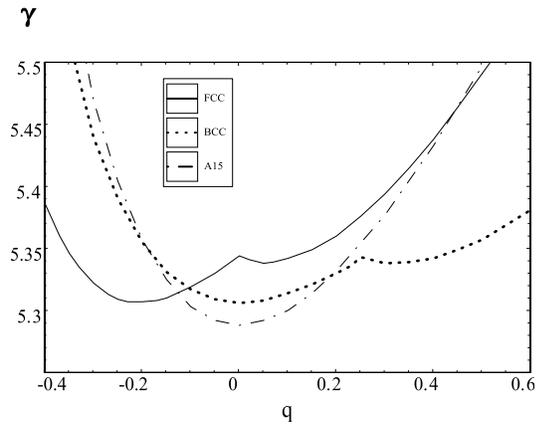}
\vspace*{5mm} \caption{The surface area parameter $\gamma$ as
function of dimensionless parameter $q$ for the elongational shear
mode. The kinks in the curves indicate a change in the topology of
the unit cell.} \label{gamma}
\end{figure}

\section{From Phases To Elasticity}

In general, the elastic energy of a crystal with cubic symmetry
has three distinct elastic constants, $K_{11},K_{12},$ and
$K_{44}$~\cite{Lubenskytext}:
\begin{eqnarray}
F_{\rm cubic}=\frac12\int {\rm d}^3x\Big[K_{11}\left(u^2_{xx}
+u^2_{yy}+u^2_{zz}\right)\nonumber\\
+K_{12}\left(u_{xx}u_{yy}+u_{xx}u_{zz}+u_{yy}u_{zz}\right)\nonumber\\
+2K_{44}\left(u^2_{xy}+u^2_{xz}+u^2_{yz}\right)\Big].
\label{elasticfree}
\end{eqnarray}
The bulk modulus $K=\frac{1}{3}\left(K_{11}+K_{12} \right)$ is an
isotropic quantity. The shear modulus, on the other hand,
generally depends on the direction of the applied strain and
ranges from deformations along the four-fold axis (elongational
shear), $\mu=K_{11}-\frac{1}{2}K_{12}$, to deformations along the
face diagonal (simple shear), $K_{44}$.  To determine the bulk
modulus as well as the three elastic constants, we calculate the
changes both in the free volume and in surface area associated
with the deformations of the lattice in question
(Fig.~\ref{deformations}).

The bulk modulus can be expressed in terms of the total free
energy of the system via:
\begin{equation}
K=V\left(\frac{\partial^2F}{\partial V^2}\right)_T
=\frac{1}{3}\left(K_{11}+K_{12}\right), \label{bulk}
\end{equation}
where $V$ is the volume of the crystal. To compute the three
elastic constants $K_{11}$, $K_{12}$, and $K_{44}$ for the fcc,
bcc, and A15 lattices in our study, we need two more measurements
which we attain from the elongational and simple shear modes.

The elongational shear mode, parametrized by the coordinate
transformation $(x,y,z)\rightarrow\left(x/\sqrt{1+q},y/\sqrt{1+q},
(1+q)z\right)$ where $q\ll1$, induces a corresponding strain
$(u_{xx},u_{yy},u_{zz})=(-q/2,-q/2,q)$. Upon substitution into
Eq.~(\ref{elasticfree}), we find:
\begin{equation}
F_{\rm{elong}}(q)=\frac{3V}{4}\left(K_{11}-\frac{1}{2}K_{12}\right)q^2
\label{K11}
\end{equation}
The volume $V$ simply results from the integration over the free
cellular volume in the definition of the elastic free energy.
Similarly, the simple shear mode, parametrized by the
coordinate transformation $(x,y,z)\rightarrow (x,y,sx+z)$,
induces the corresponding strain $u_{xz}=s/2$; all other components
of the strain tensor vanish. Analogously, we have:
\begin{equation}
F_{\rm{simple}}(s)=\frac{V}{4}K_{44}s^2.
\label{K44}
\end{equation}

The left-hand side of Eqs.~(\ref{K11}) and (\ref{K44}) can be
found within our framework.  There are two contributions from the
bulk and interfacial parts: we now find their dependence on the
deformation parameters $q$ and $s$. For example, we consider the
elastic energy of the elongational shear of the bcc lattice in
Fig.~\ref{qFbcc}. Note that the magnitude of these negative
contributions to the bulk modulus increases with increasing
density.
\begin{figure}
\vspace{2mm} \hspace{-10mm}
\includegraphics{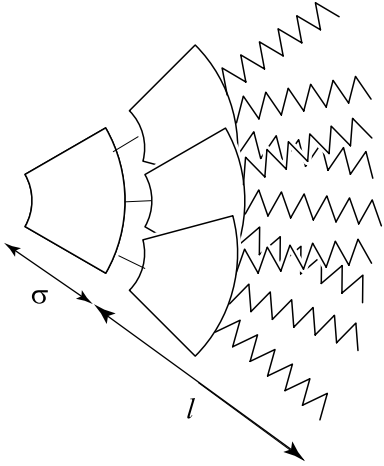}
\vspace*{5mm} \caption{The effective dimensionless "screening
length" of the dendrimers, $L=\ell/\sigma$, corresponds to the
ratio of the thickness of the soft outer part, consisting mainly
of the alkyl corona, to the radius of the impenetrable hard core
of the dendrimer molecule consisting of rigid aromatic rings. }
\label{L}
\end{figure}

The bulk elastic energies can be found via a simple numerical
scheme by deforming the unit cell and calculating the resulting
free volume of the colloidal particle. At this point, we note that
this approach reproduces the well-known shear instability of the
bcc lattice of hard spheres. Shear along a four-fold axis, called
the Bain strain, reduces the bulk free energy and leads eventually
to an fcc lattice via a continuous sequence of body-centered
tetragonal (bct) lattices. This is illustrated in Fig.~\ref{qFbcc}
where the bulk free energy of a bcc lattice is plotted as a
function of the elongational shear parameter $q$. At $q=0$, the
unit cell is cubic and the free energy reaches a local maximum.
The absolute minimum, which is located at $q=2^{1/3}-1\approx0.26$
at all densities, corresponds exactly to the fcc lattice. The
position of the local minimum at $q<0$, which describes a bct
lattice, depends on density. Its relative depth with respect to
the bcc lattice increases with density. The modulus of the
curvature near the bcc point increases with increasing reduced
density $n$.
\begin{figure}
\vspace{2mm} \hspace{-6mm}
\includegraphics{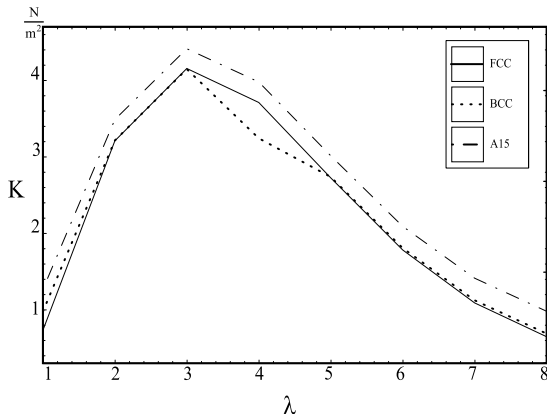}
\vspace*{10mm} \caption{The bulk modulus $K$ for the fcc, bcc, and
A15 lattices of a charged colloidal system as a function of
$\lambda=\kappa a$, where $a$ is the average interparticle
spacing. The calculations are done at density $n=0.9$ and at
dimensionless surface potential $\Phi_s=0.4$. The maximum at
$\lambda=3$ is spurious and signals the breakdown of our
approximation (see text). } \label{ChargedBulkN09}
\end{figure}

\begin{figure}
\vspace{2mm} \hspace{-6mm}
\includegraphics{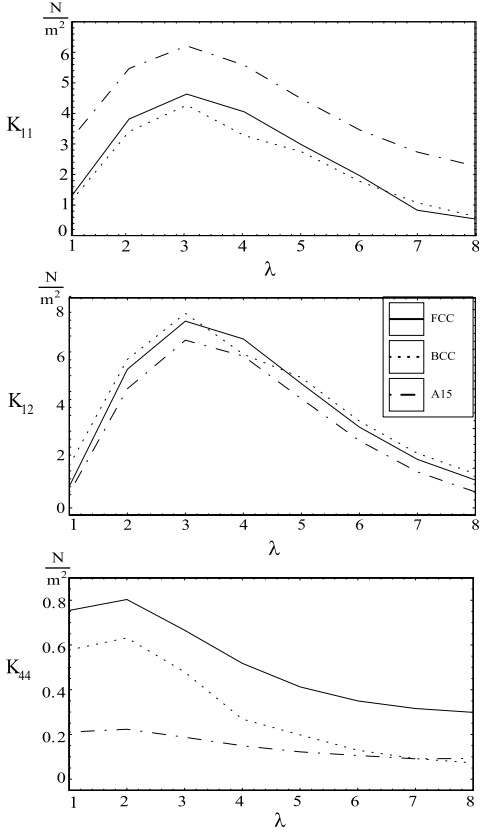}
\vspace*{3mm} \caption{The elastic constants $K_{11}$, $K_{12}$,
and $K_{44}$ for the fcc, bcc, and A15 lattices of a charged
colloidal system as a function of $\lambda=\kappa a$, at density
$n=0.9$ and at dimensionless surface potential $\Phi_s=0.4$.}
\label{ChargedK_N09}
\end{figure}

\begin{figure}
\vspace{2mm} \hspace{-6mm}
\includegraphics{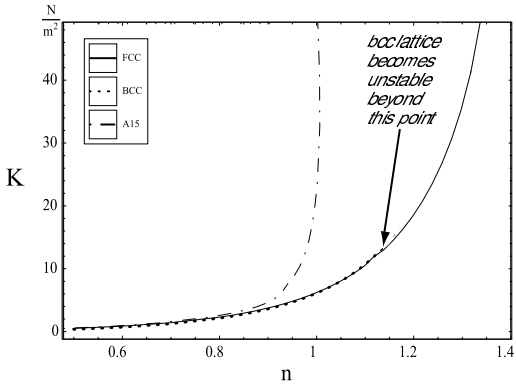}
\vspace*{3mm} \caption{The bulk modulus $K$ for the fcc, bcc, and
A15 lattices of a charged colloidal system as a function of
density $n$, at fixed screening length $\lambda=4$ and at surface
potential $\Phi_s=0.4$. The bulk modulus $K$ increases with
increasing density $n$ for all three lattices and diverges upon
close-packing.} \label{ChargedBulkL04}
\end{figure}
The computation of the interfacial elastic free energy was
facilitated by Surface Evolver~\cite{Brakke}.
Referring to Eq.~(\ref{area}), we can consider the change in
$\gamma$ with respect to the deformation parameters $q$ and $s$.
This function now encodes the change in surface area of each
Voronoi cell under the elongational and simple shear modes,
respectively. The numerical result for the elongational shear mode
is shown in Fig.~\ref{gamma}.

The stability of any particular lattice depends on both the bulk
and surface free energy terms. In the bcc lattice, the
bulk term favors the shear deformation whereas the surface term
does not. It is their relative magnitude that determines whether
the lattice is stable, and that depends on density. Since the bulk
free energy diverges at close packing, we expect that the bcc
lattice should always become unstable at high enough densities. In
the following, we only plot the elastic constants for each lattice
in the physically relevant range of parameters where the lattice
is stable.

We fit the curves shown in Fig.~\ref{gamma} with a quadratic to
extract a term proportional to the square of the strain parameter.
Upon substitution into Eqs.~(\ref{Finterfacial}) and
(\ref{Ffuzzy}), we obtain the interfacial contribution to the
elastic energy for both the charged and fuzzy colloidal systems,
respectively. We repeat the procedure for the set of bulk elastic
energy curves (Fig.~\ref{qFbcc}) and the analogous data set for
the simple shear mode. The final step requires equating the
continuum energy change from Eq.~(\ref{elasticfree}) with the
calculated energy change and solving numerically for the
coefficients in Eqs.~(\ref{bulk}), (\ref{K11}), and (\ref{K44}).

Some of our results for the case of charged colloids in a bcc
lattice have already been reported in Ref.~\cite{Kung02}. There we
found good agreement with experimental data. Here, we extend our
calculations to the fcc and A15 lattices. We also apply this
method to the case of fuzzy colloids, of which the fcc and bcc
lattices are the main candidates for the solid phase.  In each
case, we examine the behavior of the bulk modulus as well as the
three elastic constants $K_{11}$, $K_{12}$, and $K_{44}$ for the
three lattices with respect to both varying density and varying
effective ``screening" length. For charged colloids, the screening
length is the Debye screening length $\kappa^{-1}$; the analogous
concept for the fuzzy colloid is the corona thickness (Fig.~\ref{L}).

\begin{figure}
\vspace{2mm} \hspace{-6mm}
\includegraphics{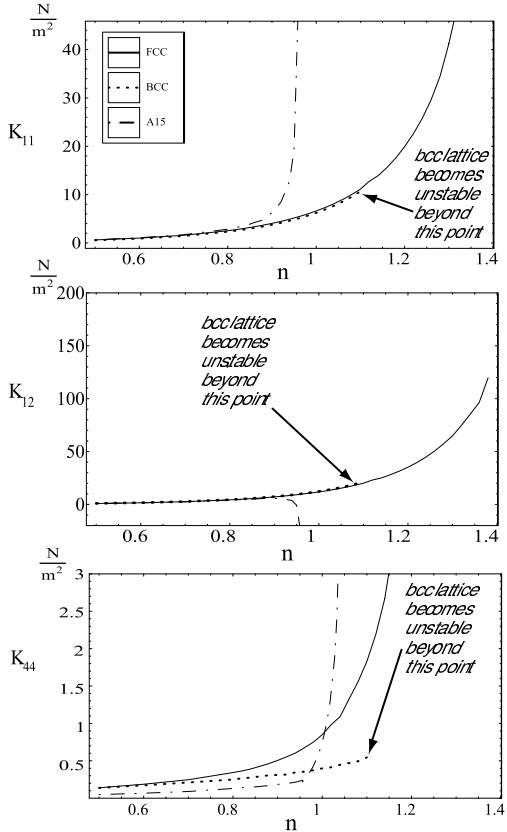}
\vspace*{5mm} \caption{The elastic constants $K_{11}$, $K_{12}$,
and $K_{44}$ for the fcc, bcc, and A15 lattices of a charged
colloidal system as a function of density $n$, at fixed
$\lambda=4$ and at surface potential $\Phi_s=0.4$. Similar to the
bulk modulus $K$, the moduli of the elastic constants diverge upon
close-packing.} \label{ChargedKL04}
\end{figure}
\begin{figure}
\vspace{2mm} \hspace{-6mm}\resizebox{3.2truein}{!}{
\includegraphics{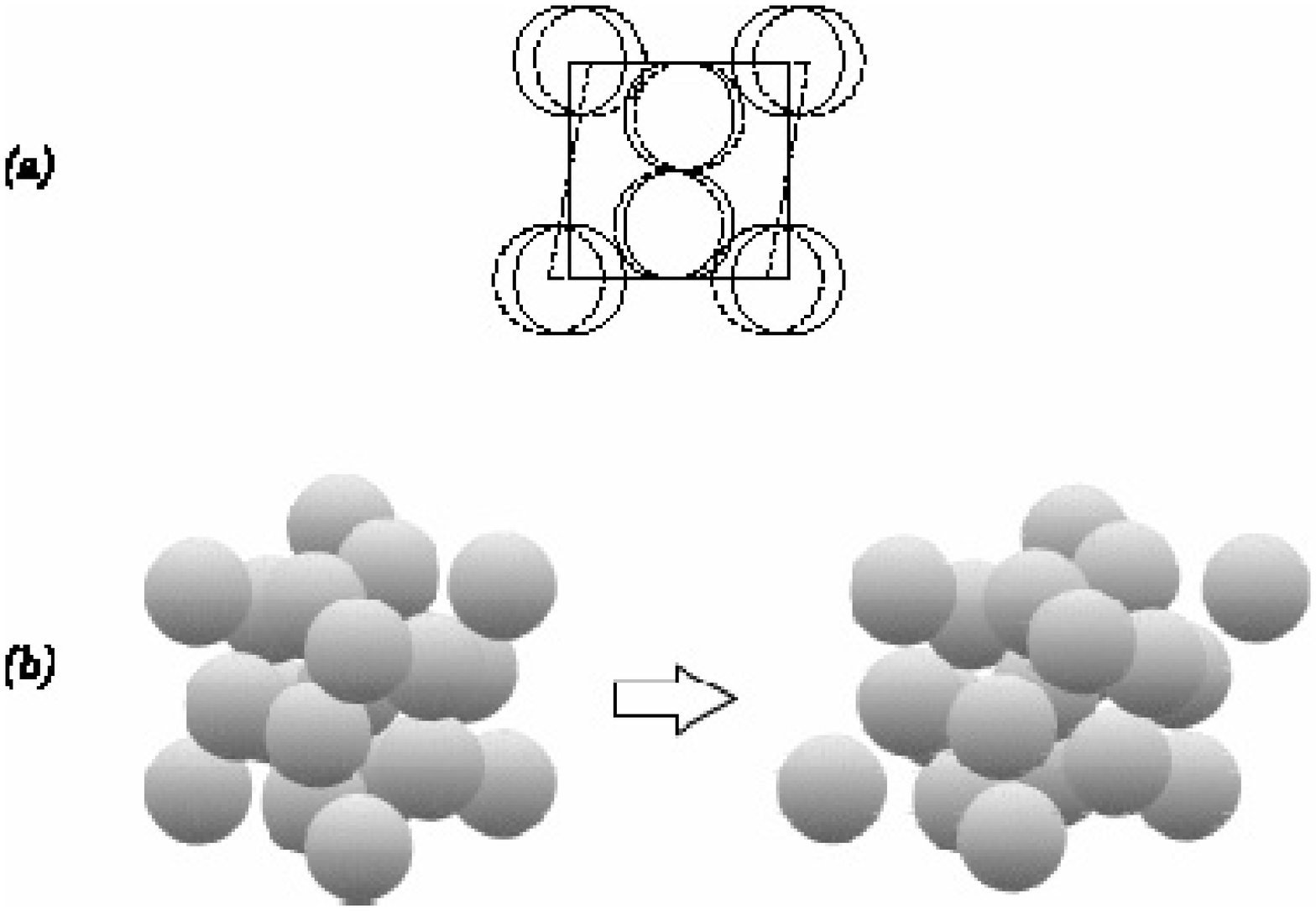}}
\vspace*{10mm} \caption{(a) Schematic of the shear instability of
the A15 lattice: the cubic arrangement of the spheres (solid line)
is unstable to shear along the face diagonal (dashed line) (b) The
cubic A15 lattice and the triclinic derivative lattice as the
entropically preferred structure at high densities.}
\label{a15shear}
\end{figure}

\subsection{Charged Colloids}

We first consider the elastic properties of charged colloids.
In what follows, the surface potential and the effective diameter
of the colloidal particles are chosen to be $\Phi_s=0.4$ and
$\sigma=910~\rm{nm}$, respectively. As a function of the
screening length at fixed density $(n=0.9)$, the bulk modulus and
three cubic elastic constants are shown in
Figs.~\ref{ChargedBulkN09} and~\ref{ChargedK_N09}, respectively.
Though the elastic constants appear to peak when the interparticle spacing
is roughly 3 screening lengths, this is an artifact of our model: as we discussed
in~\cite{Kung02}, when the interparticle spacing, $a$, is comparable to the Debye screening length,
next to nearest neighbor interactions should be included.  Since our Derjaguin-like approximation
only accounts for nearest neighbors, it is not reliable for $\lambda\equiv \kappa a$ of order 1.
Thus we expect that the peaks in the moduli mark the crossover at which the nearest neighbor approximation breaks down.
We duplicated the analysis at densities
$n=0.7$ and $0.8$, and we observe the same spurious maximum in
the plots of the elastic constants but trust our results beyond the peak.

\begin{figure}
\vspace{2mm} \hspace{-6mm}
\includegraphics{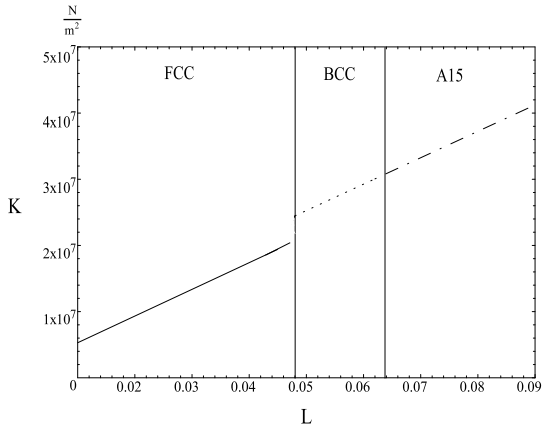}
\vspace*{3mm} \caption{The bulk modulus for the fcc, bcc, and A15
lattices of a fuzzy colloidal system as a function of "screening"
length $L=\ell/\sigma$, at fixed density $n=0.9$. As the length of
the corona increases with increasing $L$, the thermodynamically
favorable structure transits from fcc, to bcc and eventually to
the A15 lattice at high enough $L$.  The corresponding bulk
modulus simply behaves linearly in the respective regions.}
\label{FuzzyBulkN09}
\end{figure}
\begin{figure}
\vspace{2mm} \hspace{-6mm}
\includegraphics{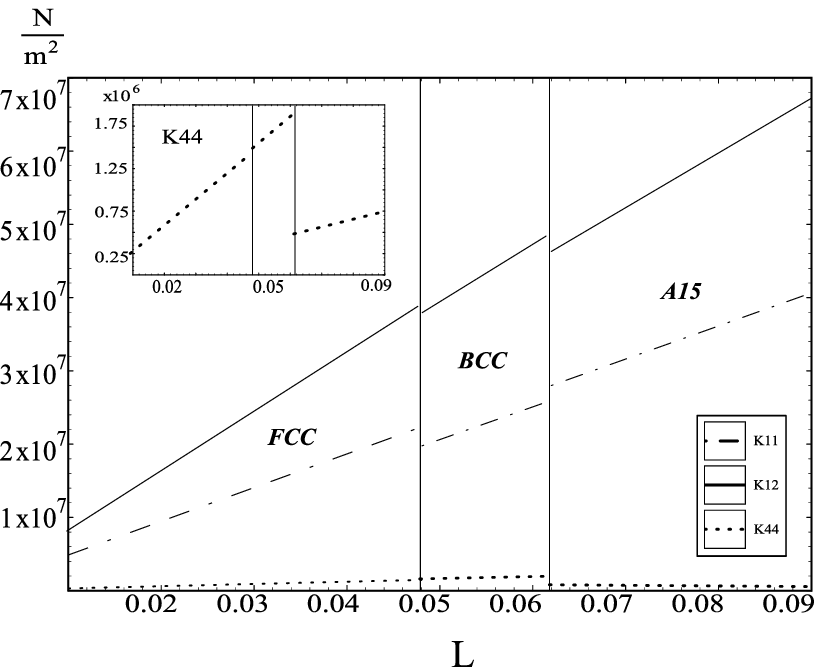}
\vspace*{3mm} \caption{The elastic constants $K_{11}$, $K_{12}$,
and $K_{44}$ for the fcc, bcc, and A15 lattices of a fuzzy
colloidal system as a function of "screening" length
$L=\ell/\sigma$, at fixed density $n=0.9$.  As the "screening"
length $L$ increases, non-close-packed structures are favored.
Similar to the bulk modulus, the three elastic constants behave
linearly with respect to $L$.} \label{FuzzyKN09}
\end{figure}

Next, we consider the elastic constants as functions of varying
density at fixed screening length so that $\lambda=\kappa a = 4$
(Figs.~\ref{ChargedBulkL04},~\ref{ChargedKL04}). Expectedly, the
bulk modulus, as well as the three elastic constants $K_{11}$,
$K_{12}$, and $K_{44}$, increase with increasing density.
Moreover, they diverge upon approaching the close packing limit.
We expect this trend to be valid for all other screening lengths
as well. The constants $K_{11}$ and $K_{12}$ which control the
elongational shear mode have similar magnitudes across different
lattices, while the simple shear constant $K_{44}$ is roughly
one-tenth of the other two constants, dramatically smaller for the
same systems. It follows that the shear moduli of the charged
colloids are considerably smaller than the bulk moduli. If we
consider the results for $K_{44}$, for instance, we see that the
modulus is largest in the fcc lattice, and significantly smaller
for the A15 lattice. This is similar to the case of hard spheres
for which it is well known that the bcc lattice is unstable with
respect to shear along the four-fold axis. It is conceivable that
for the non-closed packed lattices, there exist pockets of
unoccupied space that become available upon shear which would
allow an overall lowering of the elastic free energy. The effects
of lattice distortion are less noticeable in the bcc and A15 lattices
due to the greater availability of space for each lattice site. Physically,
this property translates into the observed fact that the bcc and
A15 lattices are much softer and much more amenable to shear
deformations. The softening of the bcc shear moduli in comparison
to the fcc ones has in fact been seen in MD studies of the fcc
lattice of copper~\cite{Kanigel01} and of the bcc lattice of
vanadium~\cite{Sorkin03}.

In the case of the A15 lattice (Fig.~\ref{a15shear}), there is an
interesting shear mode. This lattice has two distinct sites: pairs
of columnar sites are located at bisectors of the faces of the
unit cell, forming three sets of mutually perpendicular columns,
and interstitial sites are at the vertices and at the center of
the unit cell. Fig.~\ref{a15shear}a shows a side view of the A15
lattice in the close-packing limit, which is determined by the
density at which the columnar spheres touch each other---the
interstitial sites are still far from their neighbors. There is a
shear mode that can exploit this excess volume: consider the
deformation shown in Fig.~\ref{a15shear}a along the $\{110\}$
direction.  The sheared lattice (dashed line) is to be contrasted
with the cubic close-packing arrangement (solid line). The
columnar sites now no longer touch one another, and the free
energy is thus lowered in this configuration. However, as the
shear increases, the columnar sites touch the interstitial sites
and there can be no more distortion.  Moreover, since the relative
free volume of the columnar and interstitial sites depends on the
density, this instability occurs only for a range of densities
near the close-packing limit of the A15 lattice; the hard-sphere
lattice becomes unstable with respect to shear along the face
diagonal at densities greater than $n\approx 0.88$. This is
qualitatively different from the shear instability of the bcc
lattice, which is unstable at all densities.

\begin{figure}
\vspace{2mm} \hspace{-10mm}
\includegraphics{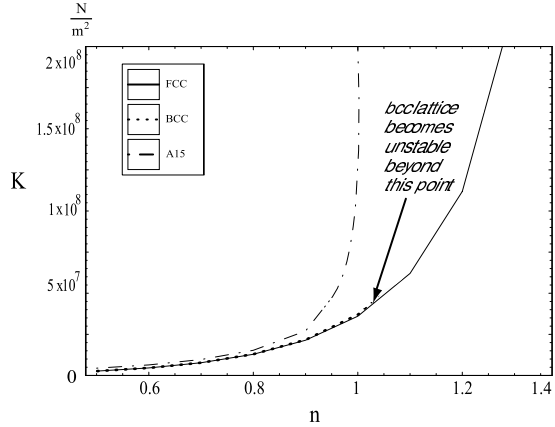}
\vspace*{13mm} \caption{The bulk modulus $K$ for the fcc, bcc, and
A15 lattices of a fuzzy colloidal system as a function of density
$n$, at "screening" length $L=\ell/\sigma=0.05$.  As expected, the
bulk modulus $K$ increases with increasing density $n$.}
\label{FuzzyBulkL005}
\end{figure}
\begin{figure} \vspace{2mm} \hspace{-6mm}
\includegraphics{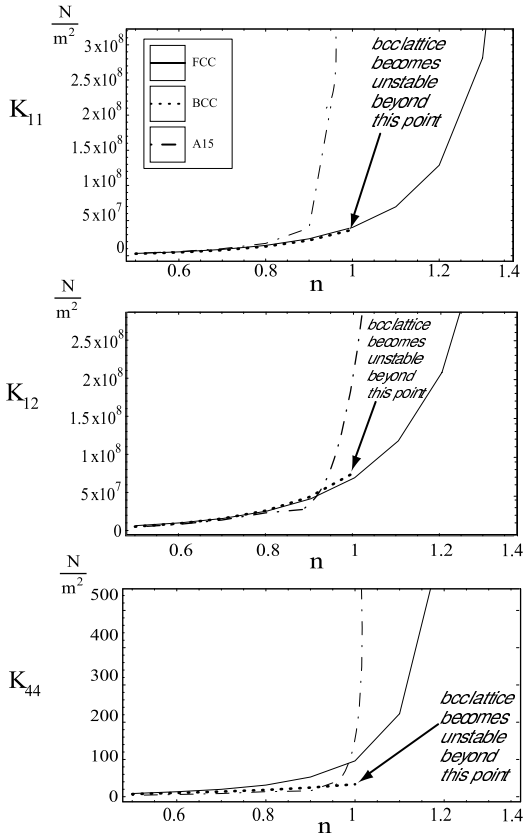}
\vspace*{3mm} \caption{The elastic constants $K_{11}$, $K_{12}$,
and $K_{44}$ for the fcc, bcc, and A15 lattices of a fuzzy
colloidal system as a function of density $n$, at "screening"
length $L=\ell/\sigma=0.05$. They increase with increasing density
$n$ much like the bulk modulus.} \label{FuzzyL005K}
\end{figure}
\begin{figure} \vspace{2mm} \hspace{-6mm}
\includegraphics{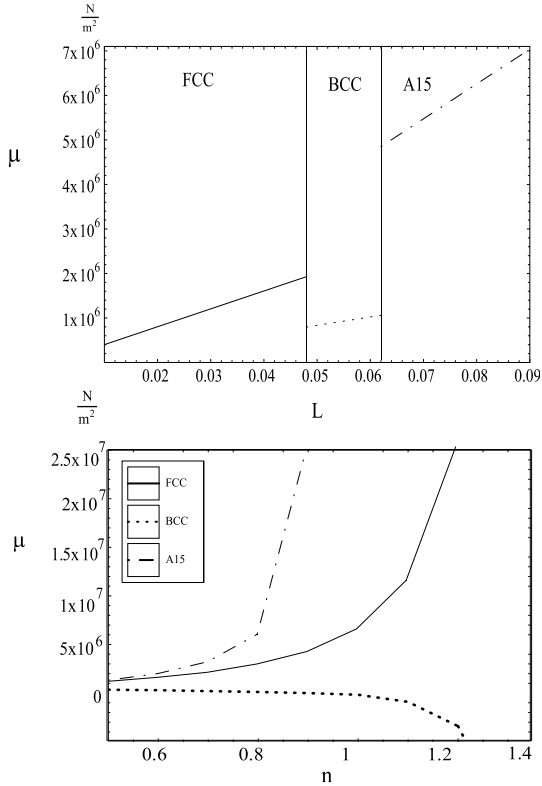}
\vspace*{6mm} \caption{The shear constants $\mu = K_{11}-K_{12}/2$
for the fcc, bcc, and A15 lattices of a fuzzy colloidal system as
a function of density $n$ and "screening" length $L=\ell/\sigma$.
They result from the elongational shear mode and represent the
lower bound of the average shear measurable in experiments.}
\label{FuzzyShear}
\end{figure}
\begin{figure} \vspace{2mm} \hspace{-6mm}
\includegraphics{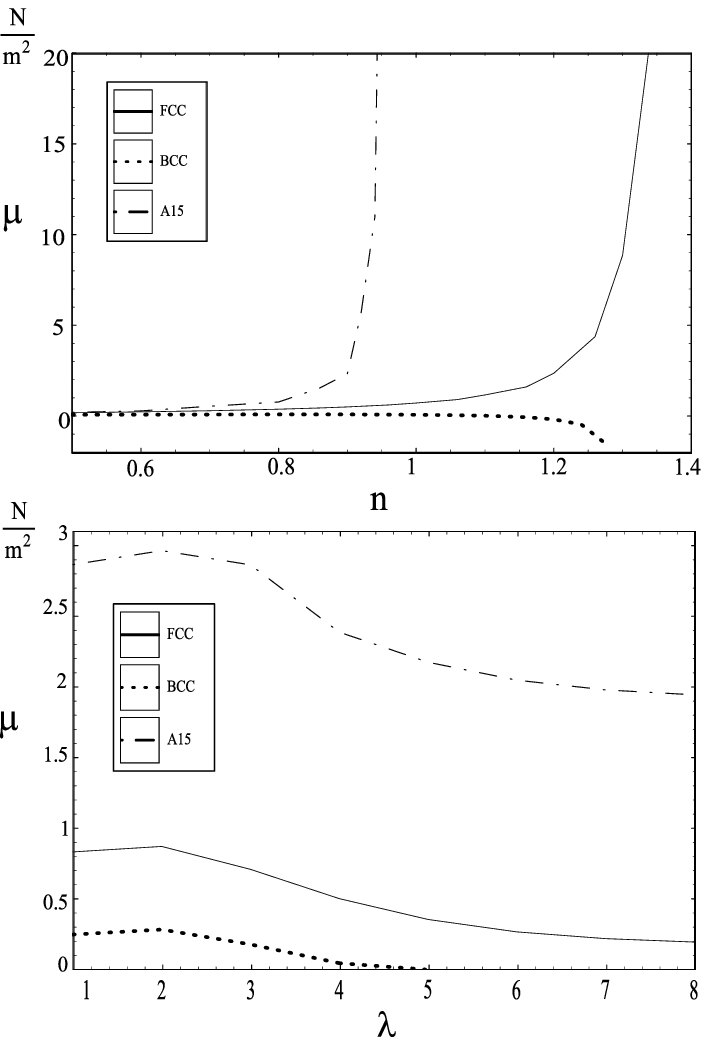}
\vspace*{6mm} \caption{The shear constants $\mu = K_{11}-K_{12}/2$
for the fcc, bcc, and A15 lattices of a charged colloidal system
as a function of density $n$ and $\lambda=\kappa a$.  They
increase with increasing density, as expected, and are the
theoretical lower bound to the range of average shear.}
\label{ChargedShear}
\end{figure}

The shear instability shown in Fig.~\ref{a15shear}a only affects a
fraction of all the sites in the unit cell and breaks the symmetry
of the cubic lattice, rendering the three columnar sites
inequivalent. Since there is no mechanism that would prefer shear
along either of the three faces of the A15 unit cell, the actual
instability experienced by the system should preserve the
equivalence of the columnar sites. Thus it should include equal
amounts of shear along all three faces, which corresponds to shear
along the body diagonal. The resulting structure belongs to the
triclinic system, since the edges of the sheared unit cell are no
longer perpendicular to one another. However, the sheared lattice
is a special triclinic lattice because the edges of the unit cell
are of equal length and the angles between them are identical. It
turns out that the angle between the edges which maximizes the
packing fraction is $\arccos(1/4)=75.5$~degrees
(Fig.~\ref{a15shear}b). This configuration gives a packing
fraction of $0.5700$, to be compared with the A15 packing fraction
of $\pi/6=0.5236$. Both this instability and the instability of
the bcc lattice can be stabilized by the soft repulsion between
the colloidal particles, encoded in the surface contribution to
the elastic free energy. In both the fuzzy and charged colloidal
systems considered in this paper, the parameters of the soft
repulsive tail are such that the A15 lattice is stable at all
densities considered.

\subsection{Fuzzy Colloids}

We now turn our attention to fuzzy colloids. As before, we
consider the effect of both varying the ``screening" length and
the density. At a fixed density of $n=0.90$, we compare the total
free energies within the high-density regime as in
Ref.~\cite{Primoz01} and find that the bcc lattice becomes more
favorable than fcc at $\ell\approx 0.043\sigma$, whereas the
bcc--A15 transition occurs at $\ell\approx 0.062\sigma$. In order
to make comparison with the previous study, we will focus on
dendrimers with $N_{0}=162$ dodecyl chains per 3rd and 4th
generation micelle with a particle size of $\sigma=4~\rm{nm}$
\cite{Balagurusamy97,Ziherl00}.  The bulk modulus and the elastic
constants are calculated for each lattice in their respective
stability region at fixed density $n=0.9$ and the results are
shown in Figs.~\ref{FuzzyBulkN09} and ~\ref{FuzzyKN09}.  Due to
the functional form of the surface energy [Eq.~(\ref{Ffuzzy})], the
linear dependence on $L=\ell/\sigma$ is preserved in the bulk
modulus and the three elastic constants.  Thus they all vary
directly with the screening length. We again observe the trend
that $K_{44}$ is smallest for the A15 lattice than for the other
lattices, while the overall magnitude of $K_{44}$ is less than
those of $K_{11}$ and $K_{12}$. Presumably, the argument presented
above for the charged colloids would apply here as well.

At fixed $L=0.05$, the plots of the bulk modulus and the elastic
constants as functions of density are shown in
Figs.~\ref{FuzzyBulkL005} and \ref{FuzzyL005K}. Except for
$K_{12}$ in the A15 lattice, they all increase with density and
diverge upon approaching close packing. The $K_{44}$ values are
also smaller than those of $K_{11}$ and $K_{12}$ across the
lattices. The elongational shear constants for the non-closed
packed structures are consistently smaller than their fcc
counterpart, in both systems of fuzzy and charged colloids.
(Figs.~\ref{FuzzyShear} and~\ref{ChargedShear}).  As with the
charged colloids, the shear modulus of the bcc lattice becomes
negative at large enough reduced density in the fuzzy colloids as
well;  only the physically relevant elastic constants have been
included in our plots.  In summary, the elastic properties of the
fuzzy colloids behave similarly to those of charged colloids, in
spite of the overall scaling of the elastic constants.  The six
orders of magnitude difference between the two systems is simply
due to the large difference in the sizes of fuzzy and charged
colloidal particles.

\section{Conclusion}

In this study, we have used the physical analogy with foams to
study the phases of crystalline lattices and their elasticity
properties. We find that while the bulk moduli of the lattices
considered are quite similar at all densities away from their
respective close-packing limits, the relative differences of their
shear moduli are much larger. The shear moduli of the
non-close-packed lattices are smaller than those of the fcc
lattice.

In many ways, the two systems considered are similar, indicating
that the details of the interparticle interaction are not
essential for all macroscopic properties of colloids.
We hope that the foam analogy used here provides an
intuitive and mathematically tangible formalism that relates
macroscopic properties of colloidal systems directly to their
microscopic constituents. Recently, such materials are becoming
increasingly important for a range of applications.
We are hopeful that our model will serve as a useful guide
for further study and synthesis, and, possibly, the engineering of
colloidal crystals of specific geometry and desired properties.
Materials with increasingly more complex crystal structures are
synthesized~\cite{Ungar03} and theoretically predicted~\cite{Likos02}
at an ever growing rate, and their understanding in terms of an
heuristic model should prove useful.

\section{Acknowledgments}

It is a pleasure to acknowledge stimulating discussions with
M. Robbins, D. Weitz and T. Witten. This work was supported by
NSF Grant DMR01-29804, the Donors of the Petroleum Research Fund,
Administered by the American Chemical Society and a gift from L. J. Bernstein.








\end{document}